# Atomically resolved scanning force studies of vicinal Si(111)

Carmen Pérez León,[1, *] Holger Drees,[1] Stefan Martin Wippermann,[2,†] Michael Marz,[1] and Regina Hoffmann-Vogel[1,‡]

[1]Karlsruhe Institute of Technology (KIT), Physikalisches Institut, Wolfgang-Gaede-Str. 1, D-76131 Karlsruhe, Germany
[2]Max-Planck-Institut für Eisenforschung GmbH, Max-Planck-Str. 1, D-40237 Düsseldorf, Germany

Well-ordered stepped semiconductor surfaces attract intense attention owing to the regular arrangements of their atomic steps that makes them perfect templates for the growth of one-dimensional systems, e.g. nanowires. Here, we report on the atomic structure of the vicinal Si(111) surface with 10° miscut investigated by a joint frequency-modulation scanning force microscopy (FM-SFM) and ab initio approach. This popular stepped surface contains 7 × 7-reconstructed terraces oriented along the Si(111) direction, separated by a stepped region. Recently, the atomic structure of this triple step based on scanning tunneling microscopy (STM) images has been subject of debate. Unlike STM, SFM atomic resolution capability arises from chemical bonding of the tip apex with the surface atoms. Thus, for surfaces with a corrugated density of states such as semiconductors, SFM provides complementary information to STM and partially removes the dependency of the topography on the electronic structure. Our FM-SFM images with unprecedented spatial resolution on steps confirm the model based on a (7 7 10) orientation of the surface and reveal structural details of this surface. Two different FM-SFM contrasts together with density functional theory calculations explain the presence of defects, buckling and filling asymmetries on the surface. Our results evidence the important role of charge transfers between adatoms, restatoms, and dimers in the stabilisation of the structure of the vicinal surface.

## I. INTRODUCTION

One-dimensional systems have been extensively studied over the last decades due to their intriguing physical properties and potential applications in nanometer-scale devices[1–7]. A bottom-up approach based on self-assembly on nanotemplates represents an attractive method for fabricating one-dimensional structures. The regular arrangement of the atomic steps of vicinal semiconductor surfaces makes them perfect templates for this approach[4,6]. In general, to exploit the advantages of the bottom-up method, it is necessary to understand the equilibrium structure of the clean vicinal surface itself[4]. Often, however, little is known about the structure of these underlying surfaces[4]. Among others, the Si(111) surface inclined by 10° towards the [$\bar{1}\bar{1}2$] direction has been widely used for the formation of ordered nanostructures[4,6,8,9]. Based on scanning tunneling microscopy (STM) images Kirakosian et al. reported that this vicinal Si(111) surface has the (557) orientation, with the period of the staircase being 5.73 nm, which correspond to 17 atomic rows[1]. Later, Teys et al.[2] proposed, also based on STM images, that this surface is oriented along the (7 7 10) rather than the initially proposed (557) [1]. Within their model, the periodically ordered steps have a height of 3 atomic layers, a width of 16 atomic rows, and a periodicity of 5.2 nm. Recently, the precise atomic arrangement of this triple step has been under debate[1,3,5,7,10].

To investigate the atomic structure of flat surfaces scanning probe microscopy techniques are widely used[11].

On conducting surfaces, STM images provide a map of the topography of the surface convoluted with its electronic structure, especially on surfaces with a corrugated density of states such as semiconductors. For these surfaces, scanning force microscopy (SFM) provides complementary information and partially removes this dependency on the electronic structure[12,13]. The atomic resolution capability of SFM arises from chemical bonding of the tip apex with the surface atoms[14]. SFM has been mainly applied to flat surfaces, because of the technical difficulties of scanning over a corrugated surface. Newly, we reported that it is possible to apply SFM and Kelvin probe force microscopy on stepped surfaces even with atomic resolution[15].

Here, we present a joint frequency modulated SFM (FM-SFM) and ab initio calculations approach to investigate the structure of the vicinal Si(111) surface at room temperature. Our atomic resolved SFM images disclose the detailed structure of the triple step with two different SFM contrasts that contribute to the understanding of the structure. To interpret this structural information, we performed density functional theory (DFT) calculations within the (16×14) surface unit cell of the Si(7 7 10) surface. The results verify the (7 7 10) orientation of the surface and reveal a number of structural details of this surface. A comprehensive analysis of the atomic arrangement of the triple step, and a rigorous comparison of our FM-SFM images with the STM data reported by Teys et al.[2,10] are also included.

## II. METHODS

Both sample preparation and experiments were carried out in an ultra-high vacuum (UHV) chamber with a base pressure of less than $3 \cdot 10^{-8}$ Pa. Stripes of low n-doped Si(111) (phosphorus, ρ = 1 − 10 Ωcm, Virginia Semiconductor) and an inclination angle of 10 ± 0.5° towards the [$\bar{1}\bar{1}2$] direction were used. The silicon stripes were cleaned in a diluted aqueous HF solution prior to loading into the UHV chamber. The sample was resistively heated by direct current with the current direction parallel to the steps on the vicinal Si(111). The surface was prepared by several short flashes to 1420 K, the last flash was proceeded by a fast ramp down to 1200 K, followed by a slower cooldown. The silicon sample was then transferred to a variable-temperature scanning force microscope (Omicron NanoTechnology GmbH, Germany) equipped with Nanosensors cantilevers (Switzerland) and a Nanonis phase-locked loop electronics (SPECS, Switzerland). All measurements were performed in the dynamic frequency modulation (FM) mode. Topographical imaging was carried out at constant frequency shift using sputter-cleaned silicon cantilevers with a force constant of 30 − 50 N/m, and a free resonance frequency of 270 − 300 kHz. Some of the cantilevers had a platinum-iridium-coated silicon tip. The long-range electrostatic interaction was minimized by applying a voltage that compensated the contact potential difference between the tip and the sample. For characterizing the FM-SFM images the normalized frequency shift ($\gamma = \Delta f \cdot k \cdot A^{3/2}/f_0$) has been used. For the discussion of the structure, the experimental drift has been compensated in Figs. 2 and 4. These images have been also slightly rotated for clarity. All measurements were performed at room temperature.

The electronic structure calculations were performed within density functional theory (DFT) and the Purdue, Burke, Ernzerhof (PBE) generalized gradient approximation[16], as implemented in the Vienna ab initio Simulation Package (VASP)[17]. The Brillouin Zone (BZ) integrations in the electronic structure calculations were done using uniform meshes, equivalent to 224 points for the (1×1) surface unit cell. The starting structures were prepared based on the experimental observation of a 16-fold lateral periodicity, the model introduced by Teys et al.[2], and variations thereof. Due to the mismatch between the 7-fold periodicity of the terraces and the 2-fold periodicity of the dimer rows, the smallest possible unit cell size parallel to the step edges that does not contain any obvious defects has a 14-fold periodicity. Thus, all calculations were performed within a (16 × 14) surface unit cell. The surface has been modeled by a slab containing four Si bilayers along the (111) direction, resulting in ∼2300 atoms per unit cell. The bottom Si bilayer was frozen at the equilibrium DFT lattice constant with the dangling bonds terminated by hydrogen. To avoid a spurious interaction between periodic images along the surface normal, a vacuum distance of 50 Å between the surface and the bottom layer of its periodic image has been employed. This vacuum distance has been used in conjunction with a dipole correction by introducing a step discontinuity inside the vacuum region which cancels out the surface dipole[18]. Forces were relaxed below a threshold of 0.01 eV/Å.

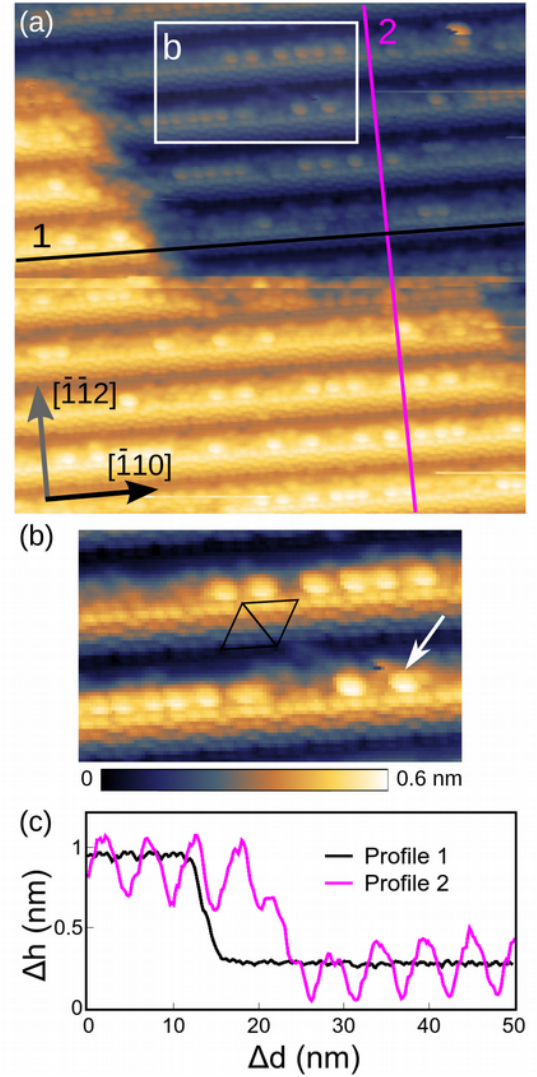

FIG. 1. FM-SFM images of the clean vicinal Si(111) surface with atomic resolution. (a) Large terraces consisting of spaced steps, separated by large step edges. Image size 50 × 50 nm². (b) Magnification 20 × 12 nm² in size of region in inset in (a). A Si(111)- 7 x 7 unit cell is indicated by two triangles. (c) Line profiles of the linecuts on (a), displaying the height and periodicity of the large and spaced steps. Imaging parameters: Δf = −16 Hz, A = 7 nm, k = 32 N/m, $f_0$ = 295 KHz, γ = −1 fN√m. Si tip.

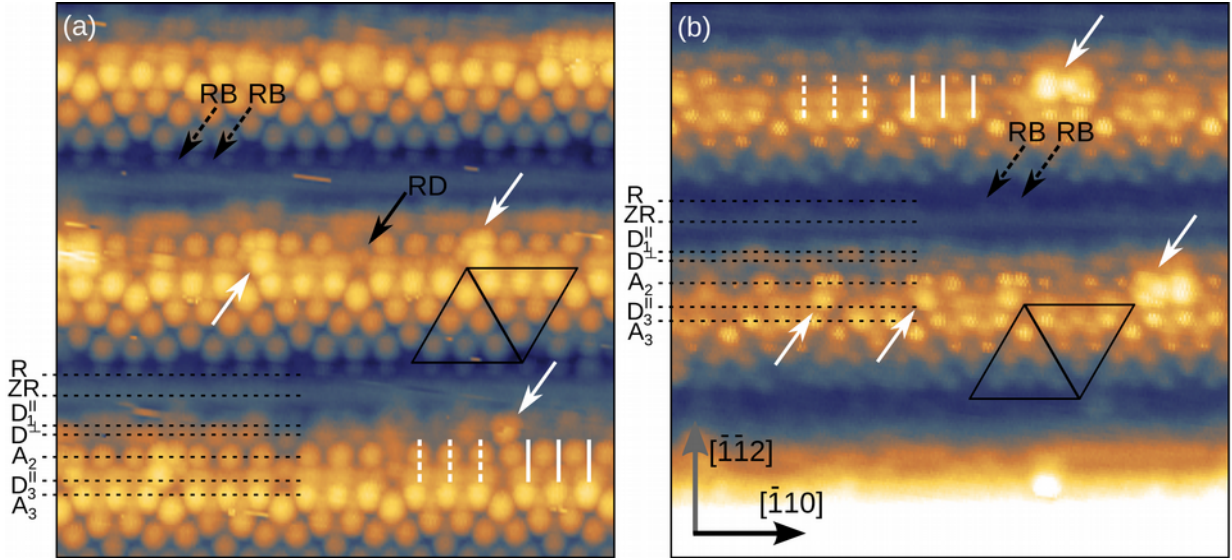

FIG. 2. Atomically resolved FM-SFM images of the vicinal Si(111) surface showing two different types of contrast: (a) normal contrast; (b) alternative contrast. A Si(111)-7 × 7 unit cell is marked with two triangles. Black dashed lines are guidelines of the triple step structure explained in detail in the text. White dashed lines and white solid lines indicate $A_2$ atoms that are located opposite to the $A_3$ atoms, or opposite to the gaps between $A_3$ atoms, respectively. Black dashed-line arrows indicate RB atoms. White arrows indicate adsorbates at step edges, and RD a row defect in an $A_2$ chain. Images size 15 × 15 nm$^2$. Imaging parameters: A = 8 nm, k = 46 N/m, $f_0$ = 272 KHz. (a) Δf = −60 Hz, γ = −7.3 fN√m. (b) Δf = −59.9 Hz, γ = −6.4 fN√m. PtIr coated Si tip.

## III. RESULTS AND DISCUSSION

Figure 1(a) shows an FM-SFM image of the silicon surface after preparation. The surface consists of large terraces, two of them can be distinguished in Fig. 1(a). Each of these large terraces contains periodically spaced steps with the height of three interplanar (111) distances and Si(111)-7 × 7-reconstructed areas. The separation of the two large terraces in Fig. 1(a) corresponds roughly to the height of two periodic steps, as shown in the profiles of Fig. 1(c). Such large terraces appear when the miscut angle of the crystal does not exactly correspond to 10°. In general, we avoid scanning with the fast axis parallel to the step edges by rotating the scan direction by 5°. Still, sometimes the tip apex becomes unstable, as it can be seen in Figure 1(a), where a jump occurred in the middle of the scan over the large step edge between the large terraces. In this image, silicon atoms are resolved in the upper and the lower large terraces. A magnification of the small region marked in Fig. 1(a) is displayed in Fig. 1(b). A unit cell of the 7 × 7-reconstructed part is indicated by two triangles. Large protrusions are observed on the step edges of the flat 7 × 7-reconstructed terraces (an example is indicated with a white arrow in Fig. 1(b)).

All SFM images in this paper are oriented like Fig. 1. This orientation implies that considering the vicinal Si(111) surface as an infinite stair, the upper part of the image correspond to the lower steps, and the lower part of the image to the higher ones. This vicinal surface is inclined towards the [$\bar{1}\bar{1}2$] direction, which corresponds to a direction parallel to the surface of the 7 × 7-reconstructed part. We do not directly show such direction in the images but a projection of it. We have decided to mark it in gray on the images axes as a guide to the eye.

In Figure 2, a close look into the atomic structure of this vicinal surface is shown. Figure 2(a) and (b) display two different FM-SFM images obtained with the same cantilever. Figure 2(a) shows the typically observed contrast, where the adatoms of the 7 × 7-reconstructed surface and the adatoms of the steps are the most protruding features. This contrast is also the most presented SFM contrast for the flat surface in the literature, and has been explained by a chemically reactive tip[19,20]. This image coincides almost one to one with the empty-states STM image[21] described in Ref. 2. Due to changes of the tip apex during scanning, in addition, the image shown in Fig. 2(b) was obtained. In this alternative contrast, the restatoms of the 7 × 7 become prominent. This kind of contrast has also been reported on the flat surface, and it has been explained by electrostatic interactions between tip and surface[22,23]. Thus, the tip change may involve the loss of one or more reactive atoms of the tip apex. Alternatively, the tip change could be dominated by a change in local contact potential difference leading to a local effective bias, and therefore, a change in the contrast. Such a bias dependence of the Si(111)-7 × 7 images is known from the literature[24]. The different contrasts highlight different features of the surface, providing supplementary information about its configuration.

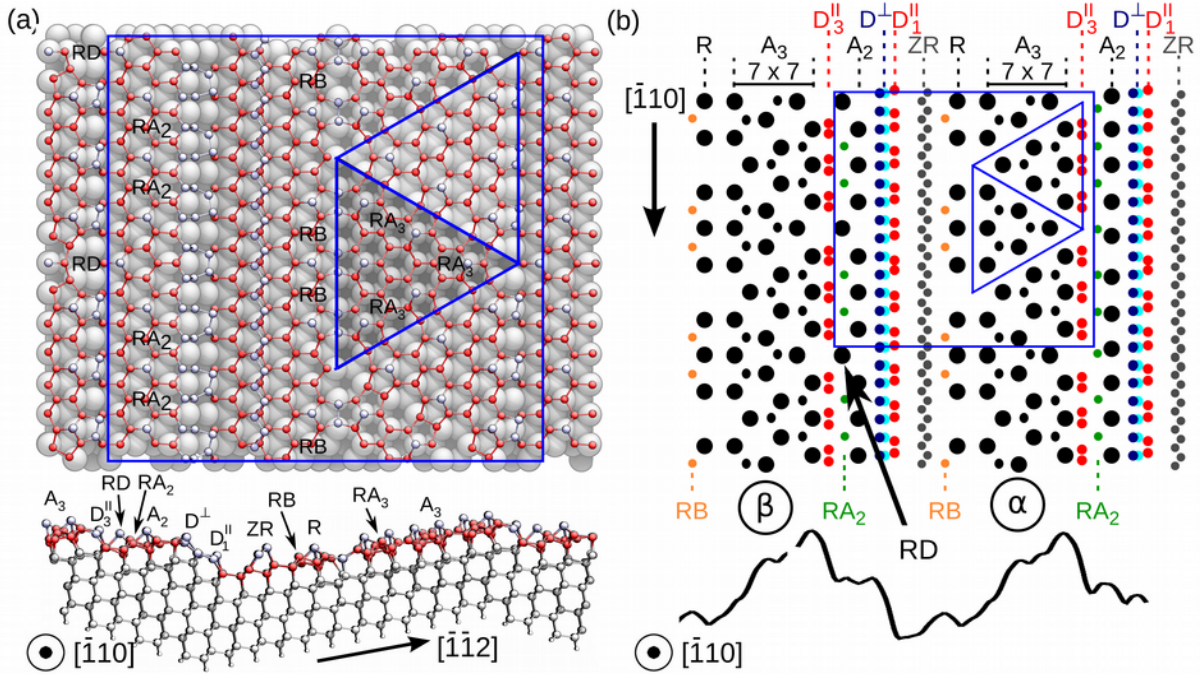

FIG. 3. (a) Top and side views of the structural model obtained from DFT calculations performed in a (16 × 14) supercell. RA$_3$, RA$_2$ and RB denote the restatoms of the surface. (b) Ball model of the top view of the vicinal Si(111) surface. The blue rectangle indicates the section shown in (a). Below SFM profile over the surface. The adatoms and restatoms of the 7 × 7 terrace, A$_2$, and R rows are plotted as black balls. Every atom of the parallel dimers ( D$^{\parallel}_3$ and D$^{\parallel}_1$ ) is plotted as a red ball. In D$^{\perp}$, the upper atoms are plotted as dark blue balls whereas the lower atoms are plotted as light blue balls. Each atom of the ZR is plotted as a grey ball. RB and RA$_2$ restatoms are plotted as orange and green balls, respectively. RD indicates a row defect in an A$_2$ chain. A Si(111)-7 × 7 unit cell is marked with two triangles.

In order to discuss the atomic arrangement of the steps in more detail, we performed DFT calculations for the Si(7 7 10) surface within the full (16 × 14) surface unit cell. Although the experimentally determined 16-fold periodicity perpendicular to the step edges strongly constrains the structural possibilities in the modeling, there are still several options how the 7 × 7 terraces and their partial counterparts are interfaced to the step edges and extra rows. Figure 3(a) shows schematic top and side views of the obtained lowest energy structure. The calculations reveal several interesting structural details in addition to the experimental data.

In the following, a detailed analysis of the atomic structure of the triple step will be presented. For this purpose the SFM images in Fig. 2(a) (normal contrast) and Fig. 2(b) (alternative contrast) will be described and compared to the resulting calculated structure and the STM data reported by Teys et al.[2,10]. We will start at the upper 7 × 7 terrace and move down along the $[\bar{1}\bar{1}2]$ direction, layer by layer. For the discussion, a ball model of the vicinal surface has been included in Fig. 3(b). In the model, the terraces will be denoted with Greek letters, and the layers with Latin letters. To enhance the features on the triple step, we have additionally applied a line subtraction fit to Fig. 2 and the resulting images are displayed in Fig. 4. For clarity we have also included a table, Table I, where the most important features of the distinct atomic species are summarized.

The terraces are formed by one cell of reconstructed 7 × 7 silicon (as indicated by two triangles in the Figures), and an additional row of atoms, R row, located at the bottom of the triple step. This additional row makes the width of the terrace slightly larger than one 7 × 7 unit cell. The R row resembles a partial 7 × 7 unfaulted half cell. Each adatom and restatom of the 7 × 7 reconstruction, and each R row atom has one dangling bond (DB) pointing upwards, i.e. perpendicular to the terrace surface, and is plotted as a black ball in the model of Fig. 3(b). As previously mentioned, in the normal contrast, Figs. 2(a) and 4(a), the higher topographic protrusions imaged are the adatoms and the R atoms, with a weak contribution of the restatoms, as in empty-states STM images. In the alternative contrast shown in Figs. 2(b) and 4(b), the adatoms are still topographically higher but they are imaged as smaller cloudy protrusions, whereas the restatoms become more noticeable. This contrast is similar to the filled-states STM image, but with the significant difference that the R row has almost disappeared in the SFM image.

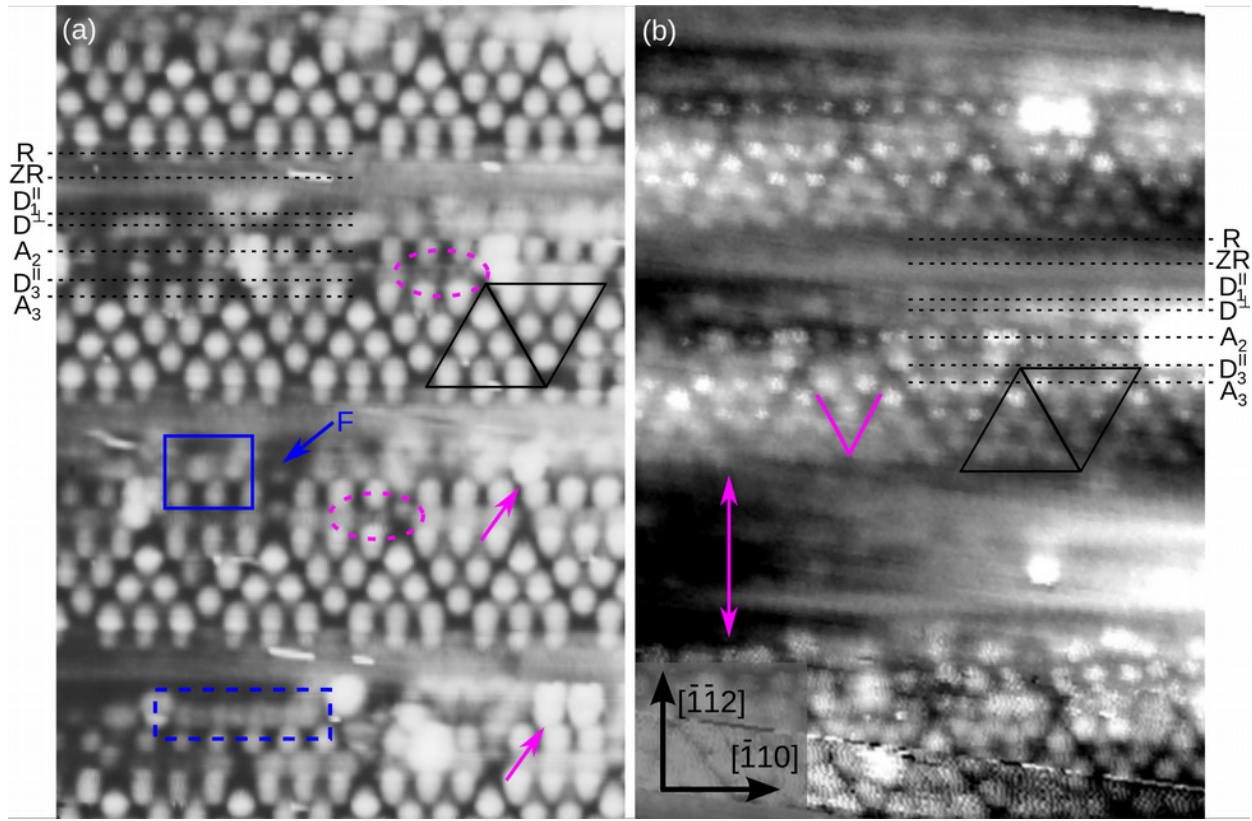

FIG. 4. A line subtraction fit was applied to the atomically resolved FM-SFM images of Fig. 2 (slighter larger area) for enhancing the features of the triple step structure. (a) Normal contrast. (b) Alternative contrast. For comparison with Fig. 2, the same unit cells are marked with two triangles. Black dashed lines are guidelines of the triple step structure. Dashed-line ovals indicate $D^{\parallel}_3$ dimers where the signal of the two atoms is split. The solid-line rectangle indicates a location where only the upper atoms of the $D^{\perp}$ dimers that are located opposite to the gaps between the $A_2$ atoms are imaged. The dashed-line rectangle indicates a location where all upper atoms of $D^{\perp}$ are imaged. Magenta arrows indicate adsorbates at step edges of the second layer ($A_2$). The blue arrow indicates a restatom in the second layer close to RD, denoted as F, that becomes visible. The V marks a defective 7 × 7 half cell. The double arrow indicates a defective triple step: the distance between the two 7 × 7 flat terraces is larger than the one usually expected for the triple step. Images size 14 × 20 nm$^2$.

On the edge of the terraces, the last row of silicon adatoms, denoted as $A_3$ (the index indicates the layer of the triple step), is accompanied by a row of parallel dimers ($D^{\parallel}_3$). Every atom of the $D^{\parallel}_3$ dimer has a DB perpendicular to the step surface, pointing slightly upwards and tilted apart from each other, and is plotted as a red ball in the model of Fig. 3(b). In general, the $D^{\parallel}_3$ atoms are seen as protrusions at the side of the $A_3$ atoms. In the normal contrast, sometimes we are able to distinguish both atoms of the dimer (see Fig. 4(a), inside the dashed-line ovals). In the alternative contrast, Figs. 2(b) and 4(b), the $D^{\parallel}_3$ dimers display a similar contrast to the restatoms. The lower layer (with index 2) is formed by a row of adatoms ($A_2$). Each $A_2$ adatom has a DB pointing upwards, i.e. perpendicular to the terrace surface, and is plotted as a black ball in the model of Fig. 3(b). The $A_2$ atoms are located either opposite to the $A_3$ atoms, as indicated with white dashed lines in Figs. 2(a) and 2(b), and depicted at step β of Fig. 3(b); or opposite to the gaps between the $A_3$ atoms as indicated with white solid lines in Figs. 2(a) and 2(b), and depicted at step α of Fig. 3(b). Often defects are observed in the $A_2$ row, in particular at the corner vacancies of the 7 × 7 surface, these are denoted as RD (row defect). At these sites there are no $D^{\parallel}_3$ dimers. Such RDs appear due to the mismatch between the 2-fold periodicity of the dimer rows and the 7-fold periodicity of the terraces. RDs are therefore, not true defects but part of the reconstruction. At the RD position, the adatom shifts towards the step, as indicated in Fig. 2(a) and Fig. 3(b). This shift makes the restatom in the second layer close to RD to become visible. This restatom is denoted as F and indicated with a blue arrow in Fig. 4(a). In general, the appearance of the adatoms of the terrace, the $A_2$ row, and R row is similar: in the normal contrast as prominent protrusions and in the alternative contrast as cloudy protrusions. Below the row of $A_2$, there are rows ascribed to perpendicular and parallel dimers ($D^{\perp}$ at the

TABLE I. Summary of the most important features of the distinct atomic species of the Si(7 7 10) surface

| Layer | Name | Art | DB orientation | Normal contrast | Altern. contrast | Model | Extra features |
|---|---|---|---|---|---|---|---|
| 3 | R | adatom | ⊥ terrace (pointing upwards) | higher topog. protrusion | almost disappeared | black | additional row |
| 3* | $RA_3$ | restatom | ⊥ terrace (pointing upwards) | weak contrib. | stronger contrib. | small black | 7 × 7 reconstruction |
| 3 | $A_3$ | adatom | ⊥ terrace (pointing upwards) | higher topog. protrusion | still higher but cloudy | black | 7 × 7 reconstruction |
| 3 | $D_3^\parallel$ | dimer | each atom: ⊥ step (slightly pointing upwards) tilted apart | diffuse protrusion sometimes both atoms resolved | similar contrast to restatoms | each atom: red | tilted row (like $D_1^\parallel$) |
| 2* | $RA_2$ | restatom | | | | green | |
| 2 | $A_2$ | adatom | ⊥ terrace (pointing upwards) | higher topog. protrusion | still higher but cloudy | black | opposite to $A_3$ or to the gap between $A_3$ |
| | (RD) | row defect | same as $A_2$ | weaker contrib. than $A_2$ | weaker contrib. than $A_2$ | like $A_2$ black | shifted towards step |
| 2* | F | restatom | | similar to RD | no data | | visible by RD & missing $D^\perp$ |
| step | $D^\perp$ | dimer | asymmetric filling upper atom: ⊥ step lower atom: saturated | many missing diffuse protrusion | many missing diffuse protrusion | dark blue / light blue | $D^\perp$-$D_1^\parallel$ buckling; $D^\perp$ between $A_2$ prefer upright orientation; $D^\perp$ opposite $A_2$ prefer flat orientation |
| 1 | $D_1^\parallel$ | dimer | each atom: ⊥ step (slightly pointing upwards) tilted apart | many missing diffuse protrusion | many missing diffuse protrusion | each atom: red | $D^\perp$-$D_1^\parallel$ buckling induces tilt in adjacent $D_1^\parallel$ |
| 1 | ZR | adatoms (zigzag) | each atom: ∥ terrace (slightly pointing upwards) away from ZR asymmetric filling | weak contrib. | weak contrib. | each atom: grey | buckled; tilted slightly towards or away from step edge |
| 0* | RB | restatom | deepest atom with DB | shoulders of ZR | shoulders of ZR | orange | |

* deeper in the layer

step, and $D_1^\parallel$ at the first layer), as indicated in Fig. 2 and Fig. 3. In the $D^\perp$ dimer, the two atoms are asymmetric: the upper atom features one DB perpendicular to the surface pointing slightly upwards (plotted as a dark blue ball in the model of Fig. 3(b)), whereas the lower atom is saturated (no DB) (plotted as a light blue ball in the model of Fig. 3(b)). Consequently, we expect to predominantly image the upper atom. In Figures 2 and 4, for both contrasts, we mainly image $D^\perp$ that are located opposite to the gaps between the $A_2$ atoms, an example is indicated with a solid-line rectangle in Fig. 4(a). In our FM-SFM images, only in the lower part of Fig. 4(a) (inside the dashed-line rectangle) all upper atoms of $D^\perp$ are imaged. Still, in the images, many dimers of the $D^\perp$ rows seem to be missing. Either these dimers are absent, or the tip is not able to interact strongly enough with them and are therefore not imaged. The low interaction with the tip may be due to their position close to the more protruding $A_2$ atoms. In the $D_1^\parallel$ dimer, as in the $D_3^\parallel$ dimer, each atom has a DB perpendicular to the step surface tilted apart from each other, and is plotted as a red ball in the model of Fig. 3(b). Also in the $D_1^\parallel$ row some dimers are missing in the SFM images. Again here, either the dimers are absent or the bond formation with the tip is sterically hindered.

The environment of the $D^\perp$ and $D_1^\parallel$ dimers with the step oriented at the (100) direction is similar to the one of the dimers on the Si(100)-(2 × 1) surface[25]. Charge transfer from the lower DB to the protruding DB causes buckling of the Si(100)-(2 × 1) dimers[26,27]. Thus, the lower DB is empty

(δ+) and the upper one is fully filled (δ−), similar to the case of a zwitterion. In our surface, however, because the $D^\perp$ has only one DB, the partner-DB is at one of the atoms of the $D^\parallel_1$ dimer. In Figure 5 the charge transfer between $D^\perp$ and $D^\parallel_1$ dimers is sketched in similarity to the one observed on the Si(100)-(2 × 1) surface. Figure 5 also displays the charge transfer between both atoms of the D 1 dimer, evidencing the interconnection of the two adjacent $D^\perp$ and the corresponding $D^\parallel_1$ dimer[10]. $D^\perp$ dimers located between two $A_2$ adatoms prefer a more upright orientation, whereas the ones directly opposite to an $A_2$ adatom assume a flatter orientation. The $D^\perp$-$D^\parallel_1$ buckling in turn induces a corresponding tilt in the directly adjacent $D^\parallel_1$ row. Interestingly, due to the mismatch between the 2-fold periodicity of the dimer rows and the 7-fold periodicity of the terraces, there can be no defect-free configuration within the buckling orientation of the $D^\perp$-$D^\parallel_1$ dimer rows. The row defects (RDs) accommodate this mismatch, making the two $D^\perp$ dimers between the $A_2$ adatoms adjacent to the RD to be both preferentially oriented in the upright configuration. However these two $D^\perp$ dimers are directly adjacent to each other, locally interrupting the 2-fold order of the buckled dimer row. In the $D^\parallel_3$ dimers there is also charge transfer between both atoms in analogy to the $D^\parallel_1$ dimers, thus the $D^\parallel_3$ dimers prefer a similar tilt angle than the $D^\parallel_1$ dimers.

Finally, in the lowest layer at the bottom of the triple step, another extra row of Si atoms is observed. This row, denoted as ZR, is at the level of the R row and shows a weak contribution to the image in Fig. 2. In Figure 4 this contribution is enhanced by the line subtraction processing, although this slightly distort the position of the row. The configuration of this row is zigzag-like as reported from STM analysis[2,10], similar to the zigzag chains observed on the Si(111)-(2 × 1) surface[28,29]. Each atom of the ZR has a DB almost parallel to the terrace slightly pointing upwards, and is plotted as a grey ball in the model of Fig. 3(b). The zigzag structure causes the DBs of consecutive atoms to point towards opposite directions, away from the ZR row. This zigzag chain is also buckled since the DBs are asymmetrically filled: the DBs of the atoms facing the R row are double filled whereas the ones of the atoms facing the $D^\parallel_1$ dimers are empty. This asymmetry has been observed in the STM images, with a shift of the bright positions between empty and filled states images. In our SFM images a similar structure of ZR is observed for both contrasts, being better resolved in the alternative contrast owing to the lack of signal from the R row. The ZR row itself is preferentially tilted slightly either towards or away from the step edge, with the tilt away from the edge being the lower energy configuration. However, even a mixed configuration within the ZR row represents a local energy minimum.

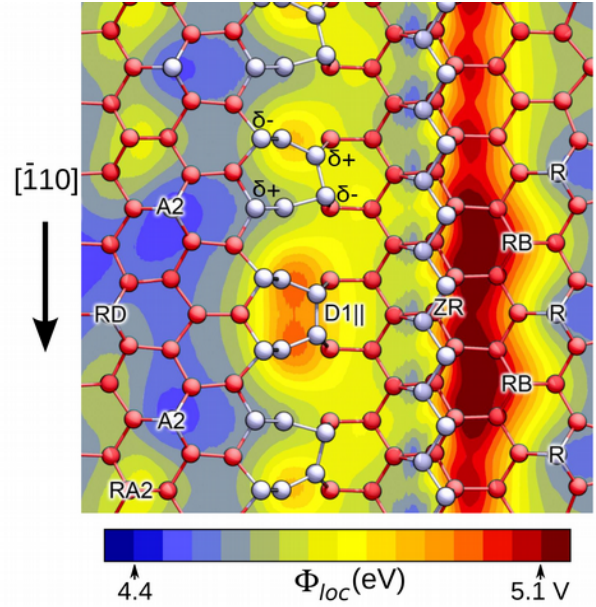

FIG. 5. Structural model of the vicinal Si(111) surface obtained from DFT calculations with the calculated local work function distribution, $\Phi_{loc}$ (for more details see Ref. 15). Charge transfer between the $D^\perp$ and $D^\parallel_1$ dimers causing buckling is indicated in similarity to the Si(100)-(2 × 1) surface.

The calculations also uncover the presence of additional restatoms. Between ZR and R, analogous to the usual restatoms of the 7 × 7 reconstruction, $RA_3$, there are restatoms denoted as RB. RB are the deepest located atoms within the reconstruction with dangling bonds. These RB restatoms are visible in both SFM contrasts as shoulders of the ZR row. In Fig. 2, RD atoms observed opposite to the gaps between the R atoms are indicated with a black dashed-line arrow. The ZR row is located at the position where the adatoms next to the RB restatoms would be if there was another 7 × 7 half cell. In addition, another restatoms are found between two RDs, and between $A_2$ adatoms and $D^\parallel_3$ dimers, denoted as $RA_2$. The area close to the corner holes at the upper part of the triple step features an absence of restatoms, due to the row defects RD passivating three potential restatom sites. In contrast, the corner holes at the bottom of the triple step (at the R row) exhibit three restatoms as the close proximity of the ZR row leaves no space for row defects within the R row analogous to the ones in the $A_2$ row (RD). RB and $RA_2$ restatoms are indicated in Fig. 3(a) and Fig. 5, and are plotted as orange and green balls in the model of Fig. 3(b), respectively.

Apart from the atomic defects reported in the description of the triple step, other structural irregularities are observed. There are several types of adsorbates at step edges, especially at the 7 × 7 terrace step edge, as mentioned above. Their contrast depends on the tip, e.g., in Fig. 2 they are better resolved than in Fig. 1. Some of these adsorbates are indicated with white arrows in Fig. 2. Also at the step edge of the second layer ($A_2$), such protrusions can be observed, as indicated with magenta arrows in Fig. 4(a). They look very similar to silicon atoms, therefore we tentatively ascribe them to additional silicon clusters produced during the preparation of the sample, while further work is needed to fully identify them[15]. Besides the Si clusters, we observe parts of the surface that are not properly reconstructed. First, on the 7 × 7 terrace in the middle of Fig. 4(b), we have marked with a V one not fully 7 × 7-reconstructed half cell. Second, in the lower part of the image, the distance between the two 7 × 7 flat terraces, marked with a double arrow in Fig. 4(b), is larger than the one usually given by the triple step.

## IV. CONCLUSIONS

Summarizing, we have investigated with a joint FM-SFM and ab initio approach the structure of the popular vicinal Si(111) surface inclined towards the [$\bar{1}\bar{1}2$] direction. We observe two different SFM contrasts: The normal atomic contrast, which is expected from the bonding of the tip apex with the dangling bonds of the atoms of the surface. And the alternative contrast, that is explained taking into account additional short-range electrostatic interactions. Our atomically resolved images with unprecedented resolution disclose the detailed structure of the triple step, and show the presence of several atomic defects, such as missing dimers and other structural irregularities. The calculations reproduce the features of the SFM image of the surface and reveal, besides the presence of different restatoms, a number of structural details on this surface: The buckling of the dimer rows at the step edges, the filling asymmetry and preferential orientation of the zigzag row, degrees of freedom in the orientation of both dimers and possibility of orientational defects. All these features originate from (partial) charge transfer between the dangling bond states.

## ACKNOWLEDGEMENTS

R.H.-V. thanks for financial support the European Research Council through the starting grant NANOCONTACTS (No. ERC 2009-Stg 239838) and the Ministry of Science, Research and Arts, Baden-Wuerttemberg, in the framework of its Brigitte-Schlieben-Lange program. S. W. acknowledges BMBF NanoMatFutur grant No. 13N12972 from the German Federal Ministry for Education and Research.

* E-mail: cperezleon.science@gmail.com
† E-mail: wippermann@mpie.de
‡ present address: Karlsruhe Institute of Technology (KIT), Institut für Angewandte Physik, Wolfgang-Gaede-Str. 1, D-76131 Karlsruhe, Germany